\documentclass[prd,preprint,nofootinbib]{revtex4}
\usepackage{amssymb}
\usepackage{epsfig}

\begin{document}

\title{ 
Dynamical solutions of warped six dimensional supergravity
}

\author{Edmund J. Copeland}
\affiliation{
 School of Physics and Astronomy, University of Nottingham, 
 University Park, Nottingham NG7 2RD, United Kingdom }
\email{Ed.Copeland@nottingham.ac.uk}

\author{Osamu Seto}
\affiliation{
 Instituto de F\'{i}sica Te\'{o}rica,
 Universidad Aut\'{o}noma de Madrid,
 Cantoblanco, 28049 Madrid, Spain }
\email{osamu.seto@uam.es}

%
\begin{abstract}
We derive a new class of exact time dependent solutions in a warped six dimensional supergravity model. Under the assumptions we make for the form of the underlying moduli fields, we show that the only consistent time dependent solutions lead to all six dimensions evolving in time, implying the eventual 
decompactification or collapse of the extra dimensions. 
We also show how the dynamics affects the quantization of the deficit angle.
\end{abstract}


\preprint{IFT-UAM/CSIC-07-30} 

\maketitle


\section{Introduction}

Six dimensional supergravity models have several interesting properties.
Salam and Sezgin obtained static solutions in which the six dimensional gauged
supergravity compactifies on a product spacetime of four dimensional Minkowski and
a two dimensional sphere, ${\cal M}_4 \times S^2$ \cite{Salam:1984cj}.
Remarkably this supergravity model admits
the supersymmetric Minkowski vacuum while many other supergravity models do not. 
The modern interpretation of this property is that the solution of this theory is compatible with the introduction of  branes into the spacetime.  
As with any massive defect, this then leads of course to the appearance of
a deficit angle in the two internal spatial dimensions, 
as a gravitational response to the tensions of the branes
 \cite{Carroll:2003db,Aghababaie:2003wz}.
The resulting geometry looks like a rugby ball solution where 
the branes are located at the north and south pole of the ball.
Gibbons, Guven and Pope (GGP) \cite{Gibbons:2003di} showed that the Salam-Sezgin vacuum is
in fact the unique one with a four dimensional maximal symmetry and
general static solutions with an axisymmetric internal space.

The observation in Ref.~\cite{Gibbons:2003di} that the four dimensional spacetime is always Minkowski
even in the presence of branes with tensions forms the basis of the interesting supersymmetric large extra dimension (SLED) scenario,
a recent  approach to solving  the cosmological constant and dark energy problems
 \cite{Aghababaie:2003wz}.
If only for this reason, such is the prize at stake, 
it makes this six dimensional supergravity interesting 
from a cosmological point of view, 
although we note that cosmology in six dimensional supergravity has previously 
been studied
in the context of Kaluza-Klein cosmology \cite{Maeda:1984gq, Maeda:1985es}.
One of the neatest aspects of the SLED model is that a 3-brane with any tension
in six dimensional spacetime induces only the corresponding deficit angle
and maintains a vanishing four dimensional cosmological constant, at least at the classical level.
This feature is often referred to as a ``self-tuning mechanism'' of the effective
four dimensional cosmological constant and would be expected to be part of 
the solution to the cosmological constant problem (although we still have to account for the affect of quantum corrections).

Although the SLED scenario has enjoyed a number of successes,
 open questions still remain. We are particularly interested in establishing  whether the self-tuning mechanism really works
in a time dependent evolving Universe (another attempt, see \cite{CDGV}). 
Previous authors have argued that the self-tuning of the four dimensional
cosmological constant does not work at least in non-supersymmetric six
dimensional Einstein Maxwell theories 
\cite{Nilles:2003km, Vinet:2004bk, Garriga:2004tq, VC}. 
The work we present here has a number of overlaps with that of Tolley et al  \cite{Tolley:2006ht},  
(and more recently with that of Kobayashi and Minamitsuji 
 \cite{Kobayashi:2007hf}) who have obtained a series of solutions 
to the dynamical system based on an elegant scaling argument. We believe that the explicit expressions presented here of exact time-dependent solutions
to the six dimensional supergravity model are given for the first time.  We begin by deriving the underlying field equations in section \ref{fieldeqn}. This is followed in section \ref{staticint} with a demonstration that it is impossible to have static internal spaces with a corresponding expanding external three dimensional space. We extend the analysis to a fully time dependent case in section \ref{timedep} and find a new class of exact solutions showing the nature of the instability and resulting evolution of the compact dimensions. 
Finally we conclude in section \ref{conc}. 

\section{The basic field equations}\label{fieldeqn}

Concentrating on the bosonic field contents of this model, 
we have the metric $g_{MN}$,
dilaton $\phi$, a $U(1)$ gauge field $A_{M}$ with field strength $F_{MN}$
and an antisymmetric tensor field $B_{MN}$ 
whose corresponding field strength is expressed as
\begin{equation}
G_{MNP} = \partial_M B_{NP} + F_{MN}A_P +{\rm cyclic \, permutations}.
\end{equation}
The lagrangian density for the bosonic sector is given by
\begin{equation}
\mathcal{L}_{\rm SUGRA} =
\frac{1}{2}R-\frac{1}{2}\partial^M \phi \partial_M \phi
-\frac{e^{-2\phi}}{12}G^{MNP}G_{MNP}-\frac{e^{-\phi}}{4}F^{MN}F_{MN}
-2 g^2 e^{\phi},
\label{Action}
\end{equation}
 where $g$ is the $U(1)$ gauge coupling
\footnote{
Note that our definition of $\phi$ and $g$ are different from those in GGP. 
They are related through 
$-2\phi_{\rm ours} = \phi_{\rm GGP}$ and $g^2_{\rm ours} = 2 g^2_{\rm GGP}$.
}.
Here $M,N$ run over all the spacetime indices and
 we work on the two - sphere of radius $r$ 
 with the six dimensional (reduced) Planck scale $M_6 = 1$.

The field equations are
\begin{eqnarray}
&& \square \phi + \frac{e^{-2\phi}}{6}G^{MNP}G_{MNP}
+\frac{e^{-\phi}}{4}F^{MN}F_{MN}-2 g^2 e^{\phi} =0, \\
&& D_M\left(e^{-2\phi} G^{MNP}\right)=0 , \\
&& D_M\left(e^{-\phi} F^{MN}\right)+e^{-2\phi}G^{MNP}F_{MP}=0 , \\
&& -R_{MN}+\partial_M \phi \partial_N \phi 
+\frac{e^{-2\phi}}{2}
\left( G_M{}^{PQ}G_{NPQ}-\frac{1}{6}G^{OPQ}G_{OPQ} g_{MN}\right) \nonumber \\
&& +e^{-\phi}
\left(F_M{}^{P}F_{NP}-\frac{1}{8}F^{PQ}F_{PQ} g_{MN}\right)+g^2e^{\phi}g_{MN}
=0,
\end{eqnarray}
and following the usual ansatz adopted for simplicity, from now on, 
we consider the case of a vanishing three form field strength
\begin{equation}
 G^{MNP} = 0 .
\end{equation}
Then, above field equations become 
\begin{eqnarray}
&& \square \phi +\frac{e^{-\phi}}{4}F^{MN}F_{MN}-2 g^2 e^{\phi} =0, \\
&& D_M\left(e^{-\phi} F^{MN}\right)=0 , \label{EOM:FMN} \\
&& -R_{MN}+\partial_M \phi \partial_N \phi 
 +e^{-\phi}\left( F_M{}^{P}F_{NP}-\frac{1}{8}F^{PQ}F_{PQ} g_{MN}\right)
+g^2e^{\phi}g_{MN} = 0.
\end{eqnarray}

The metric ansatz we adopt is 
\begin{eqnarray}
&& ds^2 = U(x^m,t)^2 ds_4 + r(t)^2 ds_2^2 , \nonumber \\
&& ds_4^2 = -dt^2+\delta_{ij}dx^i dx^j, 
 \quad ds_2^2 = \gamma_{mn}(x^m)dx^m dx^n, \label{MetricAnsatz}
\end{eqnarray}
where $i, j$ run over the usual three spatial indices, $m, n$ run over the extra two spatial indices and $\gamma_{mn}(x^m)$ is an arbitrary two-dimensional metric.
With this metric, the two form field strength takes the form of 
\begin{eqnarray}
&& F_{\mu\nu} = F_{\mu m} = 0 , \nonumber \\
&& F_{mn} = F(t,x^m) \epsilon_{mn} ,\label{FMN}
\end{eqnarray}
 with $\epsilon_{mn}$ being the anti-symmetric tensor.
Here, the use of Greek indices denote 
the four spacetime coordinates (i.e. $\mu = (0, i)$).
With the metric ansatz Eqn.~(\ref{MetricAnsatz}),
we write each component of  the Einstein equations, 
$(0-0), (0-m), (i-j), (m-n)$ respectively as 
\begin{eqnarray}
%
&& 3 \left(\frac{\partial_0 U}{U}\right)_{,0} 
 +2\left[\left(\frac{\partial_0 r}{r}\right)_{,0}
 -\frac{\partial_0 r}{r}\frac{\partial_0 U}{U} 
 +\left(\frac{\partial_0 r}{r}\right)^2\right] 
 +\partial_0\phi\partial_0\phi \nonumber \\
&& 
 -\frac{2}{r^2}\gamma^{mn}\partial_m U \partial_n U  
 -\frac{1}{2 r^2}D_m\left(\gamma^{mn} \partial_n U^2\right) 
 +\frac{e^{-\phi}}{8}F^{PQ}F_{PQ} U^2 -g^2e^{\phi}U^2 = 0 , \\
%
&& -\left(\frac{\partial_m U}{U}\right)_{,0}
 +4\left(\frac{\partial_0 U}{U}\right)_{,m} 
 -4\frac{\partial_m U}{U}\frac{\partial_0 r}{r}
 +\partial_0\phi\partial_m \phi =0 , \\
%
&& -\left(\frac{\partial_0 U}{U}\right)_{,0} 
 -2\left(\frac{\partial_0 U}{U}\right)^2 
 -2\frac{\partial_0 U}{U}\frac{\partial_0 r}{r}
  \nonumber \\
&&  
 +2\frac{\partial_m U}{U}\frac{\partial_n U}{U}
 \frac{U^2}{r^2}\gamma^{mn} 
 +\frac{1}{2 r^2}D_m(\gamma^{mn}\partial_n U^2) 
 -\frac{e^{-\phi}}{8}F^{PQ}F_{PQ} U^2 +g^2e^{\phi}U^2 = 0 , \\
%
&& -\left[ \left(\frac{\partial_0 r^2}{2 U^2}\right)_{,0}
 +4\frac{\partial_0 U}{U}\frac{\partial_0 r}{r}\frac{r^2}{U^2} 
 \right] \gamma_{mn} +4D_n\left(\frac{\partial_m U}{U}\right)
 +4\frac{\partial_m U}{U}\frac{\partial_n U}{U} - R^{m'}{}_{mm'n} \nonumber \\
&&  +\partial_m\phi\partial_n\phi 
 +e^{-\phi}\left( F_m{}^{P}F_{Pn}-\frac{1}{8}F^{PQ}F_{PQ} g_{mn}\right)
 +g^2e^{\phi}g_{mn} = 0 .
\end{eqnarray} 

\subsection{The static Gibbons, Guven and Pope solution}

Before discussing the time dependent solutions of this system,
we recall the derivation of the static solution originally obtained  by
Gibbons, Guven and Pope (GGP) \cite{Gibbons:2003di}.
When we take $r(t)=1$, $U(x^m,t)=W(x^m)$ and $\phi = \phi(x^m)$, 
 the Einstein equations and the equation of motion for the dilaton $\phi$ reduce to
\begin{eqnarray}
&& 
 -\frac{1}{4 W^4}D_m (\gamma^{mn}\partial_n W^4) 
 +\frac{e^{-\phi}}{8}F^{PQ}F_{PQ}  -g^2e^{\phi} = 0 ,
\label{GGP:Einstein} \\
&& -4D_n\left(\frac{\partial_m W}{W}\right)
 -4\frac{\partial_m W}{W}\frac{\partial_n W}{W} + R^{m'}{}_{mm'n} \nonumber \\
&&  -\partial_m\phi\partial_n\phi 
 -e^{\phi}\left( F_m{}^{P}F_{nP}-\frac{1}{8}F^{PQ}F_{PQ} \gamma_{mn}\right)
 -g^2e^{\phi}\gamma_{mn} = 0 ,
\end{eqnarray} 
and
\begin{equation}
\frac{1}{W^4}D_m\left(\gamma^{mn}W^4\partial_n \frac{\phi}{2}\right) 
+\frac{e^{-\phi}}{8}F^{MN}F_{MN}- g^2 e^{\phi} =0 ,
\label{GGP:EOM}
\end{equation}
leading to the GGP solution: 
\begin{equation}
 \phi = -\frac{1}{2} \ln W^4. \label{GGPsol:phi-W}
\end{equation}
The equation of motion (\ref{EOM:FMN})  for $F_{mn}$ leads to 
 the solution to Eqn.~(\ref{FMN})
\begin{equation}
 F = -\frac{q}{2} e^{\phi}W^{-4} = -\frac{q}{2} W^{-6},
\end{equation}
where we have used Eq.~(\ref{GGPsol:phi-W}). Note that $q$ can be interpreted as a magnetic charge.

It will prove useful to recall the
 explicit solution for $\phi$ and $\gamma_{mn}$ as presented by GGP
  \cite{Gibbons:2003di, Gibbons:1987ps}
\begin{eqnarray}
&& \gamma_{mn} = diag (\gamma_{rr}, \gamma_{\psi,\psi}), \nonumber \\
&& (\gamma_{rr}, \gamma_{\psi\psi}) = \left( \frac{e^{-\phi}}{ f_0^2},
 \frac{e^{-\phi} r^2}{f_1^2} \right), \\
&& e^{2\phi} = \frac{f_0}{f_1} , 
\end{eqnarray}
with 
\begin{eqnarray}
&& f_0 \equiv 1+\frac{r^2}{r^2_0}, \quad f_1 = 1+\frac{r^2}{r^2_1}, 
 \label{f0f1} \\
&& r_0^2 =\frac{1}{g^2},  \quad  r_1^2 = \frac{8}{q^2} \label{r0r1} .
\end{eqnarray} 

In the following sections we begin to explore the dynamical equations by allowing the scale factors and the fields to become time dependent.

\section{Only one evolving space is not a solution}\label{staticint}

\subsection{Time dependent U}

Ideally what we want to obtain is a solution
 which describes the expansion of our three space dimensions with
a static extra dimensional space.
As a first step towards obtaining it, we make the ansatz of a static internal space $r=1$, and a static dilaton $\phi = \phi(x^m)$. 
This combination makes sense in that the relation $r^2 \propto e^{-\phi}$ has previously been obtained in \cite{Aghababaie:2003wz} hence the dilaton would be static if $r$ is static. 
In fact, in the following subsection, we will show that
allowing for a time dependence of $\phi$ does not improve the possibility of obtaining static $r$ solutions.  
The field equations are reduced to
\begin{eqnarray}
&& \frac{3}{U^2} \left(\frac{\partial_0 U}{U}\right)_{,0} 
 -\frac{1}{4 U^4}D_m (\gamma^{mn}\partial_n U^4)
 +\frac{e^{-\phi}}{8}F^{PQ}F_{PQ} -g^2e^{\phi} = 0 ,
 \label{W:00}\\
%
&& \left(\frac{\partial_m U}{U}\right)_{,0}
 -4\left(\frac{\partial_0 U}{U}\right)_{,m}  =0 ,
 \label{W:0m}\\
%
&& -\frac{1}{U^2}\left(\frac{\partial_0 U}{U}\right)_{,0}
 -\left(\frac{\partial_0 U}{U}\right)^2 \frac{2}{U^2} 
 +\frac{1}{4 U^4}D_m (\gamma^{mn}\partial_n U^4)
  -\frac{e^{-\phi}}{8}F^{PQ}F_{PQ}  +g^2e^{\phi} = 0 ,
 \label{W:ij} \\
%
&& 4D_n\left(\frac{\partial_m U}{U}\right)
 +4\frac{\partial_m U}{U}\frac{\partial_n U}{U} - R^{m'}{}_{mm'n} \nonumber \\
&&  +\partial_m\phi\partial_n\phi 
 +e^{-\phi}\left( F_m{}^{P}F_{Pn}-\frac{1}{8}F^{PQ}F_{PQ} \gamma_{mn}\right)
 +g^2e^{\phi}\gamma_{mn} = 0 ,
 \label{W:mn} \\
&&
\frac{1}{U^4}D_m\left(\gamma^{mn}U^4\partial_n \frac{\phi}{2}\right) 
 + \frac{e^{-\phi}}{8}F^{MN}F_{MN}- g^2 e^{\phi} =0. 
 \label{W:phi}
\end{eqnarray} 
Although the equations look intractible, 
we can make progress by noting that because
\begin{eqnarray}
\left(\frac{\partial_m U}{U}\right)_{,0}
= \frac{\partial_0\partial_m U}{U}-\frac{\partial_0 U \partial_m U}{U^2}
=\left(\frac{\partial_0 U}{U}\right)_{,m} ,
\end{eqnarray}
 then from Eq.~(\ref{W:0m}),
\begin{eqnarray}
\left(\frac{\partial_m U}{U}\right)_{,0}
=\left(\frac{\partial_0 U}{U}\right)_{,m} =0 .
\end{eqnarray}
The general solution of $U$ is therefore given by
 $\ln U = \ln a(t)+ \ln W(x^m)$ where 
 $a(t)$ and $W(x^m)$ are integration functions.
Thus, we find that $U$ has to take the separable form of $U = a(t)W(x^m)$. The field equations~(\ref{W:00}) and (\ref{W:ij}) can then be reduced to
\begin{eqnarray}
&& C(x^m)
 -\frac{1}{4 W^4}D_m (\gamma^{mn}\partial_n W^4)
 +\frac{e^{-\phi}}{8}F^{PQ}F_{PQ} -g^2e^{\phi} = 0 , \\
&& \frac{3}{U^2} \left(\frac{\partial_0 a}{a}\right)_{,0} 
  = \frac{1}{U^2}\left(\frac{\partial_0 a}{a}\right)_{,0} 
 + \frac{2}{U^2}\left(\frac{\partial_0 a}{a}\right)^2 
  = C(x^m),
\end{eqnarray}
which after some algebra leads to
\begin{eqnarray}
&& a(t) = \frac{a_0}{t - t_0} , \\
&& C(x^m) = \frac{3}{a_0^2 W(x^m)^2} ,
\end{eqnarray}
with $a_0$ and $t_0$ being integration constants.

However, from Eqs.~(\ref{W:00}) and (\ref{W:phi}), we also know that 
\begin{eqnarray}
D_m\left(\gamma^{mn}U^4 \partial_n (\ln U^4+ 2\phi )\right) -4 U^4 C(x^m) =0, 
\label{W:00-phi}
\end{eqnarray}
 which is conflict, for any non-vanishing $U$ on the internal space, 
 with the fact that the extra two dimensional space is compact.
In other words, 
if we integrate both sides of Eq.~(\ref{W:00-phi}) over the compact extra space,
 we see that the first term on the left hand side vanishes as it is a total derivative while the second term does not. Hence, there is an inconsistency and so we conclude there is no static solution for the compact space with this ansatz for $U$.

There is a caveat to this argument. We have implicitly assumed that the extra space is smooth.
However, if we allow the extra space to be singular then it is possible that, 
de-Sitter type solutions may be obtained~\cite{Tolley:2005nu}.

\subsection{Time dependent $U$ and $\phi$, but static $r$.}

We now allow for a time-dependent dilaton.
The field equations are
\begin{eqnarray}
%
&& 3 \left(\frac{\partial_0 U}{U}\right)_{,0} +\partial_0\phi\partial_0\phi 
 -\frac{1}{4 U^2}D_m\left(\gamma^{mn} \partial_n U^4\right)
 +\frac{e^{-\phi}}{8}F^{PQ}F_{PQ} U^2 -g^2e^{\phi}U^2 = 0 , \\
%
&& -\left(\frac{\partial_m U}{U}\right)_{,0}
 +4\left(\frac{\partial_0 U}{U}\right)_{,m} 
 +\partial_0\phi\partial_m \phi =0 , \\
%
&& \left(\frac{\partial_0 U}{U}\right)_{,0} 
 +2\left(\frac{\partial_0 U}{U}\right)^2 
 -\frac{1}{4 U^2}D_m\left(\gamma^{mn} \partial_n U^4\right)
 +\frac{e^{-\phi}}{8}F^{PQ}F_{PQ} U^2 -g^2e^{\phi}U^2 = 0 , \\
%
&& 4D_n\left(\frac{\partial_m U}{U}\right)
 +4\frac{\partial_m U}{U}\frac{\partial_n U}{U} - R^{m'}{}_{mm'n} \nonumber \\
&&  +\partial_m\phi\partial_n\phi 
 +e^{-\phi}\left( F_m{}^{P}F_{Pn}-\frac{1}{8}F^{PQ}F_{PQ} g_{mn}\right)
 +g^2e^{\phi}g_{mn} = 0 ,\\
&& -\frac{1}{U^4}\partial_0\left( U^2 \partial_0 \frac{\phi}{2}\right)
 +\frac{1}{U^4}D_m\left( U^4 \gamma^{mn} \partial_n \frac{\phi}{2}\right)
 +\frac{e^{-\phi}}{8}F^{MN}F_{MN}- g^2 e^{\phi} =0 .
\end{eqnarray} 
If we again assume the form $U=a(t)W(x^m)$, 
 then the $(0-m)$ component of the Einstein equation implies $\phi=\phi(t)$. 
In addition, from the $\phi$ equation of motion, using the solution for the flux
 $F \propto U(t,x^m)^{-4}$ , we find that $W=1$  
 because each term has a different dependence on $W$ and 
 the $\phi$ equation of motion can not be satisfied if $U$ depends on $x^m$.
Hence we must have $U=a(t)$ and moreover since the spacetime no longer has a non-trivial warp factor  $W(x^m)$, it can not be warped. Given the above result, the equations of motion now can be written as:
\begin{eqnarray}
%
&& 3 \left(\frac{\partial_0 a}{a}\right)_{,0} +\partial_0\phi\partial_0\phi 
 +\frac{e^{-\phi}}{8}F^{PQ}F_{PQ} a^2 -g^2e^{\phi}a^2 = 0 , \label{Up:00}\\
%
&& \left(\frac{\partial_0 a}{a}\right)_{,0} 
 +2\left(\frac{\partial_0 a}{a}\right)^2 
 +\frac{e^{-\phi}}{8}F^{PQ}F_{PQ} a^2 -g^2e^{\phi}a^2 = 0 , \\
&& -\frac{1}{a^2}\partial_0\left( a^2 \partial_0 \frac{\phi}{2}\right)
 +\frac{e^{-\phi}}{8}F^{MN}F_{MN}a^2- g^2 e^{\phi}a^2 =0,\\
%
&& \frac{3}{4}e^{-\phi}F^{PQ}F_{PQ} + 2g^2e^{\phi} 
 = g^{mn} R^{m'}{}_{mm'n} (x^m) \equiv R_c(= {\rm const}). \label{Up:mn}
\end{eqnarray} 
Notice that the term $g^{mn} R^{m'}{}_{mm'n} $ 
could in principle be a function of $x^m$, but in this case it is not allowed  
 by Eq.~(\ref{Up:mn}) as the left hand side  depends only on $t$.
This fact means that the compact extra space must be 
a constant curvature two dimensional sphere. 
Here there is no way to introduce branes which induce a deficit angle and 
deform a sphere with a constant curvature into a rugby ball shape.
Therefore, we can see that this ansatz, namely varying $U$ and $\phi$ with static $r$ 
can not lead to satisfactory solutions.

\subsection{Time dependent $r$ and $\phi$, but static $U$.}

Finally, let us try to obtain a solution of the static three space with 
a dynamical extra dimension.
If we take $r=r(t)$ and $\phi = \phi(t, x^m)$, but $U=W(x^m)$, 
then Einstein's equations and the equation of motion for $\phi$ are
\begin{eqnarray}
%
&& 
 2\left[\left(\frac{\partial_0 r}{r}\right)_{,0}
 +\left(\frac{\partial_0 r}{r}\right)^2\right] +\partial_0\phi\partial_0\phi 
 -\frac{D_m(\gamma^{mn}\partial_n W^4)}{4 r^2 W^4}
 +\frac{e^{-\phi}}{8}F^{PQ}F_{PQ} -g^2e^{\phi} = 0 , \label{rp:00}\\
%
&& -4\frac{\partial_m W}{W}\frac{\partial_0 r}{r} 
 +\partial_0\phi\partial_m \phi =0 ,
 \label{rp:0m} \\
%
&& 
 \frac{D_m(\gamma^{mn}\partial_n W^4) }{4 r^2 W^4}
 -\frac{e^{-\phi}}{8}F^{PQ}F_{PQ} +g^2e^{\phi} = 0 , \label{rp:ij} \\
%
&& 4D_n\left(\frac{\partial_m W}{W}\right)
 +4\frac{\partial_m W}{W}\frac{\partial_n W}{W} - R^{m'}{}_{mm'n}  
 -\frac{1}{2U^2}\left(\partial_0 r^2\right)_{,0}\gamma_{mn} \nonumber \\
&&  +\partial_m\phi\partial_n\phi 
 +e^{-\phi}\left( F_m{}^{P}F_{Pn}-\frac{1}{8}F^{PQ}F_{PQ} g_{mn}\right)
 +g^2e^{\phi}g_{mn} = 0 , \label{rp:mn} \\
&& -\frac{1}{W^2r^2}\partial_0\left(r^2 \partial_0 \frac{\phi}{2}\right)
 +\frac{1}{W^4r^2}D_m\left( W^4 \gamma^{mn} \partial_n \frac{\phi}{2}\right) +\frac{e^{-\phi}}{8}F^{MN}F_{MN}- g^2 e^{\phi} =0. \label{rp:phi}
\end{eqnarray} 
Now provided
 that $\phi(t,x^m)$ can be decomposed as $\phi(t,x^m) = \phi(t) + \phi(x^m) $
 and $F$ depends only on $x^m$, then 
 Eqs.~(\ref{rp:00}) and (\ref{rp:ij}) can be reduced to
\begin{eqnarray}
&& r(t)^2 e^{\phi(t)} =1 \label{rp:rExp} ,  \label{rp:r-phi}  \\
&&  \left(\frac{\partial_0 r}{r}\right)_{,0}
 + \left(\frac{\partial_0 r}{r}\right)^2 
 +\frac{1}{2}\partial_0\phi\partial_0\phi =0 , \\
&& \frac{D_m(\gamma^{mn}\partial_n W^4) }{4 W^4}
 -\frac{e^{-\phi(x^m)}}{8}F^{PQ}F_{PQ} +g^2e^{\phi(x^m)} = 0 .
\end{eqnarray}
Unfortunately, the solution of these equations are not compatible with
\begin{equation}
\partial_0 r^2 = 0 ,
\end{equation}
which can be  obtained from Eq.~(\ref{rp:phi}) using Eq.~(\ref{rp:r-phi}).
Thus, we once again see that there is no consistent solution with this ansatz.

\section{Time dependent solutions with dymanical $r, \, U$ and $\phi$}\label{timedep}

Having tried  unsuccessfully to obtain static solutions for $r$ and $U$, we now look for dynamical solutions where all the key fields $r, U$ and $\phi$ are time dependent. We again make a series of ansatz, in this case $\phi(t,x^m) = \phi(t) + \phi(x^m) $, and assume the separable form of $U = a(t)W(x^m)$. The Einstein equations and the equation of motion for $\phi$ are:
\begin{eqnarray}
&& \frac{3}{U^2} \left(\frac{\partial_0 a}{a}\right)_{,0} 
 +\frac{2}{U^2}\left[\left(\frac{\partial_0 r}{r}\right)_{,0}
 -\frac{\partial_0 r}{r}\frac{\partial_0 a}{a} 
 +\left(\frac{\partial_0 r}{r}\right)^2\right] 
 +\frac{\partial_0\phi\partial_0\phi}{U^2} \nonumber \\
&&  
 -\frac{1}{4 r^2 W^4}D_m\left(\gamma^{mn} \partial_n W^4\right)  
 +\frac{e^{-\phi}}{8}F^{PQ}F_{PQ} -g^2e^{\phi} = 0 ,
\label{CosmoEOM:R00} \\
&& -4 \frac{\partial_m W}{W}\frac{\partial_0 r}{r}
 +\partial_0\phi\partial_m \phi =0 ,
\label{CosmoEOM:R0m} \\
&& \frac{1}{U^2}\left(\frac{\partial_0 a}{a}\right)_{,0} 
 +\frac{2}{U^2}\left(\frac{\partial_0 a}{a}\right)^2 
 +\frac{2}{U^2}\frac{\partial_0 a}{a}\frac{\partial_0 r}{r}  
\nonumber \\
 &&  
 -\frac{1}{4 r^2 W^4}D_m(\gamma^{mn}\partial_n W^4)
 +\frac{e^{-\phi}}{8}F^{PQ}F_{PQ} -g^2e^{\phi} = 0 ,
\label{CosmoEOM:Rij} \\
%
&& 4 D_n\left(\frac{\partial_m W}{W}\right)
 +4 \frac{\partial_m W}{W}\frac{\partial_n W}{W} - R^{m'}{}_{mm'n} 
 -\left[ \left(\frac{\partial_0 r^2}{2 W^2}\right)_{,0}
 +4 \frac{\partial_0 a}{a}\frac{\partial_0 r}{r}\frac{r^2}{U^2} 
 \right] \gamma_{mn} \nonumber \\
&&  +\partial_m\phi\partial_n\phi 
 +e^{-\phi}\left( F_m{}^{P}F_{Pn}-\frac{1}{8}F^{PQ}F_{PQ} g_{mn}\right)
 +g^2e^{\phi}g_{mn} = 0 ,
\label{CosmoEOM:Rmn} \\
&& -\frac{1}{U^4r^2}\partial_0\left(r^2 U^2 \partial_0 \frac{\phi}{2}\right)
 +\frac{1}{W^4r^2}D_m\left( W^4 \gamma^{mn} \partial_n \frac{\phi}{2}\right)
 +\frac{e^{-\phi}}{8}F^{MN}F_{MN}- g^2 e^{\phi} =0 .
\label{CosmoEOM:phi}
\end{eqnarray}
Under the additional ansatz that $e^{\phi(t)} r^2 =1$ 
(motivated by the observation that $r^2 \propto e^{-\phi}$
 \cite{Aghababaie:2003wz}), which is equivalent to 
\begin{equation}
 \partial_0\phi = -2\frac{\partial_0 r}{r}\,,
\label{CosmoEOM:DerTime:phi-r} 
\end{equation}
 we see that Eq.~(\ref{CosmoEOM:R0m}), leads to
\begin{equation}
\partial_m\phi = -2\frac{\partial_m W}{W} ,
\label{CosmoEOM:DerX:phi-W} 
\end{equation}
 as in the GGP solution. If we further assume that the field strength $F$ is static and only depends on $x^m$,  then the field equations  Eqs.~(\ref{CosmoEOM:R00}), (\ref{CosmoEOM:Rij}) and 
 (\ref{CosmoEOM:phi}) coupled with Eq.~(\ref{CosmoEOM:DerX:phi-W}) 
 can be rewritten as the following differential equation 
 which depends only on $x^m$,
\begin{equation}
 C(x^m) -\frac{D_m(\gamma^{mn}\partial_n W^4) }{4 W^4}
 +\frac{e^{-\phi} r^2}{8}F^{PQ}F_{PQ} -g^2e^{\phi}r^2 = 0 ,
\label{Cconstraint}
\end{equation}
 where, $C(x^m)$ is given by
\begin{eqnarray}
 C(x^m) &=& \frac{r^2}{U^2} \left[3 \left(\frac{\partial_0 a}{a}\right)_{,0} 
 +2\left(\frac{\partial_0 r}{r}\right)_{,0}
 - 2\frac{\partial_0 r}{r}\frac{\partial_0 a}{a} 
 +2\left(\frac{\partial_0 r}{r}\right)^2 
 + \partial_0\phi\partial_0\phi \right] \nonumber \\
 &=& \frac{r^2}{U^2}\left[ \left(\frac{\partial_0 a}{a}\right)_{,0} 
 +2 \left(\frac{\partial_0 a}{a}\right)^2 
 +2 \frac{\partial_0 a}{a}\frac{\partial_0 r}{r} \right] \nonumber \\
 &=& -\frac{1}{U^4}\partial_0\left(r^2 U^2 \partial_0 \frac{\phi}{2}\right) ,
\label{CosmoEOM:definitionK}
\end{eqnarray}
 each equality in Eq.~(\ref{CosmoEOM:definitionK}) arising  from 
Eqs.~(\ref{CosmoEOM:R00}), (\ref{CosmoEOM:Rij}) and (\ref{CosmoEOM:phi}), 
respectively.

\subsection{Power law solutions for $r$ and $a$}

Eqns.~(\ref{Cconstraint}) and (\ref{CosmoEOM:definitionK}) still look very difficult to solve directly from first principles, and so instead we will try to obtain solutions by assuming the form of $r$ and $a$,  and looking for self-consistency in the solutions. As a first attempt we assume power law behaviour for them, namely:
\begin{eqnarray}
a \propto t^{n}, \quad r \propto t^{n_r} .
\label{Pol:rW}
\end{eqnarray}
The three equalities in Eq.~(\ref{CosmoEOM:definitionK})
coupled with Eq.~(\ref{CosmoEOM:DerTime:phi-r}) now become
\begin{eqnarray}
 C(x^m) &=& \frac{t^{2(n_r-n-1)}}{W^2}(-3n-2n_r-2n_r n+6n_r^2) \nonumber \\
 &=& \frac{t^{2(n_r-n-1)}}{W^2}n_r(2n+2n_r-1)  \nonumber \\
 &=& \frac{t^{2(n_r-n-1)}}{W^2}n(2n+2n_r-1) .
 \label{CosmoEOM:CinPower}
\end{eqnarray}
There are two possible ways in which $C(x^m)$ can be a function of only $x^m$,
as required by Eq.~(\ref{Cconstraint}). The first is if the time dependent prefactor vanishes which corresponds to $n_r-n-1=0$. The second way is if the right hand side of each of the terms vanish, which corresponds to the brackets vanishing in Eq.~(\ref{CosmoEOM:CinPower}).
The former is precisely the structure found by Tolley et al~\cite{Tolley:2006ht} based on a scaling argument for the scale factors. However, the metric ansatz of $g_{\mu\nu}$ in ~\cite{Tolley:2006ht} 
is slightly different to ours (\ref{MetricAnsatz}).
In particular it follows that the condition $n_r-n-1=0$ is not a solution in our case, because it can not satisfy all three equalities in Eqs.~(\ref{CosmoEOM:CinPower}). In fact 
we determine the values of $n$ and $n_s$ in Eqs.~(\ref{CosmoEOM:CinPower}) by equating  the coefficients :
\begin{equation}
 -3n-2n_r-2n_r n+6n_r^2 = n_r(2n+2n_r-1) = n(2n+2n_r-1).
\end{equation}
This has the non trivial solution
\begin{eqnarray}
n = \frac{2\pm\sqrt{3}}{4}, \quad  n_r = \mp\frac{\sqrt{3}}{4}.
\end{eqnarray}
This in turn gives $C(x^m)=0 $ which is consistent with the above discussions 
and is compatible with Eq.~(\ref{CosmoEOM:Rmn}) too, 
because the solution satisfies
\begin{equation}
 \left(\frac{\partial_0 r^2}{2 a^2}\right)_{,0}
 +4 \frac{\partial_0 a}{a}\frac{\partial_0 r}{r}\frac{r^2}{a^2} = 0 .
\end{equation}
Notice that in this case, we obtain identical solutions for  $F$, $\phi$ and $W$ as found in the GGP solution. This is as expected, since the $x^m$ dependent part of the field equations are identical to that of the GGP solution. In this sense we have obtained the time dependent version of the GGP solution.

From the metric ansatz Eqn.~(\ref{MetricAnsatz}) it follows that the time $t$ is actually the  conformal time in the usual sense. The ``cosmic time'' $\tau$ can therefore be defined as $d \tau \propto t^{n}dt$, from which we obtain
\begin{eqnarray}
&& ds^2 = W(x^m)^2 [-d\tau^2+a(\tau)^2 \delta_{ij}dx^i dx^j] 
 + r(\tau)^2 ds_2^2, \nonumber \\
&& a(\tau) \propto \tau^{n/(n+1)} , \quad r(\tau) \propto \tau^{n_r/(n+1)},
\end{eqnarray}
in terms of the cosmic time. 

\subsection{Exponential solutions  for $r$ and $a$}

The next obvious step is to assume an exponential form
\begin{eqnarray}
 a(t) = e^{h t}, \quad r(t) = e^{h_r t}, 
\end{eqnarray}
 where $h$ and $h_r$ are constants. 
In this case, Eq.~(\ref{CosmoEOM:definitionK})
 with Eq.~(\ref{CosmoEOM:DerTime:phi-r}) leads to
\begin{eqnarray}
 C(x^m) &=& \frac{e^{2(h_r-h)t}}{W^2}2h_r(-h + 3 h_r) \nonumber \\
 &=& \frac{e^{2(h_r-h)t}}{W^2}2h_r(h+h_r)  \nonumber \\
 &=& \frac{e^{2(h_r-h)t}}{W^2}2h(h+h_r) .
 \label{CosmoEOM:CinExp}
\end{eqnarray}
which now has a non trivial solution
$h=h_r$. Then, $C(x^m)$ is given by
\begin{eqnarray}
 C(x^m) = \frac{4h^2}{W(x^m)^2}.
\end{eqnarray}
Thus, we obtain the equations of motion of the $x^m$ dependent part 
of the fields to be
\begin{eqnarray}
&& \frac{4h^2}{W(x^m)^2} -\frac{D_m(\gamma^{mn}\partial_n W^4) }{4 W^4}
 +\frac{e^{-\phi(x^m)}}{8}F^{PQ}F_{PQ} -g^2e^{\phi(x^m)} = 0 , \\
&& 
 \frac{4 h^2}{W(x^m)^2} \gamma_{mn}
 -\partial_m\phi\partial_n\phi 
 -e^{-\phi(x^m)}\left( F_m{}^{P}F_{Pn}-\frac{1}{8}F^{PQ}F_{PQ} \gamma_{mn}\right)
 -g^2e^{\phi(x^m)}\gamma_{mn} = 0 \label{XMeqn}.
\end{eqnarray}
Something significant can now be seen. Recall that we have equation~(\ref{CosmoEOM:DerX:phi-W}), relating $\phi$ and $W(x^m)$. Given the solution we have just obtained, we see that in Eqn.~(\ref{XMeqn}), by 
introducing $\tilde{g}^2 \equiv g^2-4h^2$, then a new solution to the system is obtained  which looks identical to the original $x^m$ part of the  GGP solution but with our redefined gauge coupling $\tilde{g}^2$ replacing the original $g^2$ coupling.
Obviously, the $h \rightarrow 0$ limit corresponds to the original static GGP solution.
Hence, we have obtained the explicit expression of the solution
including  the $x^m$ dependent part.
However, notice that since $h$ is just a constant  it could in principle take any value.
In particular, for $4h^2 > g^2$, 
corresponding to a negative $\tilde{g}^2$,  we find that the $x^m$ dependent part of the solution has only differs slightly from that obtained in GGP. 
We show this and give the actual solution for the case of vanishing
 and negative $\tilde{g}^2$ in Appendix B.
The line element of this solution with such a nonvanishing $h$ 
is rewritten as
\begin{eqnarray}
ds^2 = W(x^m)^2 [-d\tau^2+(h \tau)^2\delta_{ij}dx^i dx^j] + (h \tau)^2 ds_2^2 ,
\end{eqnarray}
in terms of the cosmic time.
This solution is the same as that found in Ref.~\cite{Tolley:2006ht}, however,
 the $x^m$ dependence of the solution was not explicitely solved for there.
Here, we have shown that it is same as that of the GGP solution.

\section{Conclusions}\label{conc} 

We have derived a new class of exact time dependent solutions in a six dimensional gauged supergravity compactified on a two dimensional axisymmetric space.
Under the assumption of a separable form of $U$ 
we showed that there is no solution expressing the either 
an expanding four dimensional universe with
a static internal space or visa versa. 
Exact solutions we obtained involved all the 
dimensions either expanding or contracting which means the eventual 
decompactification or collapse of the extra dimension, 
indicating an instability of Salam-Sezgin,
${\rm (Minkowski)}_4 \times S^2$, spacetime for the case with the absence of the maximal symmetry in the four dimensional spacetime. 

In the above analysis, we did not include into the action brane terms such as 
\begin{equation}
 S_{\rm brane}= \sum_i \int d^4x \, T_i =
 \sum_i \int d^6x \, T_i \delta^{(2)}(x^m-x^m_i )\,, 
\end{equation}
where $T_i$ is the tension of the `i-th' brane and 
$x^m_i$ denotes the position of the brane in the internal space.
However, we can easily introduce 
such brane terms, their affect being to induce the deficit 
angle in the internal space. 
The topological condition for the gauge field $A_M$ is the same as 
we previously obtained for the static solutions, because 
the solution of the gauge field strength $F_{mn}$ is unchanged in the presence of the branes.
As is discussed in \cite{Gibbons:2003di}, 
for the case of $r_0^2 \neq r_1^2$ in Eq.~(\ref{f0f1}),
while one pole can be smooth, the other has a deficit angle
\begin{equation}
\frac{\delta}{2\pi}=1-\frac{r_1^2}{r_0^2}.
\label{defangle}
\end{equation}
Combining Eq.~(\ref{r0r1}) and 
the topological condition, the Dirac quantization condition, 
for the gauge field $A_M$ becomes  \cite{Gibbons:2003di, Aghababaie:2003ar},
\begin{equation}
\frac{4g}{q} =N,
\end{equation}
leading to the quantized deficit angle 
\begin{equation}
\frac{\delta}{2\pi}=1-N^2 ,
\end{equation}
which was previously obtained for the static solution 
with $N$ being an integer \cite{Gibbons:2003di}.
However, for the new solution in Section IV B, Eq.~(\ref{defangle}) is rewritten 
as
\begin{equation}
\frac{\delta}{2\pi}=1-\frac{8\tilde{g}^2}{q^2} ,
\label{defpos}
\end{equation}
 for a positive $\tilde{g}^2$,
\begin{equation}
\frac{\delta}{2\pi}=1 ,
\label{defzero}
\end{equation}
 for a vanishing $\tilde{g}^2$ and,
\begin{equation}
\frac{\delta}{2\pi}=1-\frac{8(-\tilde{g}^2)}{q^2} ,
\label{defneg}
\end{equation}
for a negative $\tilde{g}^2$.
Hence the deficit angle is given as
\begin{equation}
\frac{\delta}{2\pi}=1-N^2 \left|1-4\frac{h^2}{g^2}\right|.
\end{equation}
This implies that the interval of the quantized deficit angle becomes narrow 
for $2h \approx g$ in the time-dependent solution. 
It would be interesting to investigate the consequence of this new deficit angle.

%
\section*{Acknowledgments}

We would like to thank Cliff Burgess and Andrew Tolley for very helpful detailed correspondence.
The work of O.S. is in part supported by PPARC, 
the MEC project FPA 2004-02015 and 
the Comunidad de Madrid project HEPHACOS (No.~P-ESP-00346).

%
\appendix

\section{conventions and geometrical quantities}

Here, we note our conventions and several geometrical quantities.

Christoffel symbols 
\begin{equation}
 \Gamma^M_{NP} = \frac{1}{2}g^{MQ}(g_{QN,P}+g_{QP,N}-g_{NP,Q}) .
\end{equation}

Riemann tensor 
\begin{equation}
 R^M{}_{NOP} = \Gamma^M_{NP,O}-\Gamma^M_{NO,P}
 +\Gamma^M_{QO}\Gamma^Q_{NP}-\Gamma^M_{QP}\Gamma^Q_{NO} .
\end{equation}
With this definition, the sign in front of the Einstein term in the action 
is a plus.

Metric
\begin{equation}
ds^2 = U^2(t,x^m)(-dt^2+ \delta_{ij}dx^i dx^j)+r(t)^2\gamma(x^m)_{mn}dx^m dx^n,
\end{equation}
where $i, j, ...,$ run over the usual three-spatial dimensions and
 $m, n, ...,$ run over the extra spatial dimensions.

Christoffel symbols 
\[
\begin{array}{ll}
 \Gamma^0_{00} = \frac{1}{2}\frac{\partial_0 U^2}{U^2} 
& \Gamma^0_{0m} = \frac{1}{2}\frac{\partial_m U^2}{U^2} \\
 \Gamma^0_{ij} = \frac{1}{2}\frac{\partial_0 U^2}{U^2}\delta_{ij} 
& \Gamma^0_{mn} = \frac{1}{2}\frac{\partial_0 r^2}{U^2}\gamma_{mn} \\
 \Gamma^i_{0j} = \frac{1}{2}\frac{\partial_0 U^2}{U^2}\delta^i{}_j 
& \Gamma^i_{jm} = \frac{1}{2}\frac{\partial_m U^2}{U^2}\delta^i{}_j \\
 \Gamma^m_{00} = \frac{1}{2}\frac{\partial_n U^2}{r^2}\gamma^{mn} 
& \Gamma^m_{0n} = \frac{1}{2}\frac{\partial_0 r^2}{r^2}\delta^m{}_n \\
 \Gamma^m_{ij} = -\frac{1}{2}\frac{\partial_n U^2}{r^2}\gamma^{mn}\delta_{ij} 
& \Gamma^m_{np} = \frac{1}{2}\gamma^{mq}
 (\gamma_{qn,p}+\gamma_{qp,n}-\gamma_{np,q}) \\
{\rm others} = 0 &
\end{array}
\] 

Ricci tensors
\begin{eqnarray}
R_{00} &=& -\delta^i{}_i \left(\frac{\partial_0 U}{U}\right)_{,0} 
 + \frac{1}{r^2}\gamma^{mn}\partial_m U \partial_n U 
 (\delta^i{}_i -1) \nonumber \\
&&  +\left[-\left(\frac{\partial_0 r}{r}\right)_{,0}
 +\frac{\partial_0 r}{r}\frac{\partial_0 U}{U} 
 -\left(\frac{\partial_0 r}{r}\right)^2\right]\delta^n{}_n
 +\frac{D_m\left(\gamma^{mn} \partial_n U^2\right)}{2 r^2} , \\
R_{0m} &=& \left(\frac{\partial_m U}{U}\right)_{,0}
 -(\delta^i{}_i+1)\left(\frac{\partial_0 U}{U}\right)_{,m} 
 +\frac{\partial_m U}{U}\frac{\partial_0 r}{r}(\delta^i{}_i+\delta^n{}_n-1) , \\
R_{ij} &=& \left[ \left(\frac{\partial_0 U}{U}\right)_{,0} 
 +\left(\frac{\partial_0 U}{U}\right)^2 (\delta^k{}_k-1)
 -\frac{\partial_m U}{U}\frac{\partial_n U}{U}
 \frac{U^2}{r^2}\gamma^{mn}(\delta^k{}_k-1) \right. \nonumber \\
&&  \left. +\frac{\partial_0 U}{U}\frac{\partial_0 r}{r}\delta^m{}_m
 -\frac{D_m(\gamma^{mn}\partial_n U^2)}{2 r^2}
 \right]\delta_{ij} , \\
R_{mn} &=& -(\delta^i{}_i+1)D_n\left(\frac{\partial_m U}{U}\right)
 -(\delta^i{}_i+1)\frac{\partial_m U}{U}\frac{\partial_n U}{U} + R^{m'}{}_{mm'n} \nonumber \\
&& 
 +\left[ \left(\frac{\partial_0 r^2}{2 U^2} \right)_{,0}
 +\frac{\partial_0 U}{U}\frac{\partial_0 r}{r}\frac{r^2}{U^2}(1+\delta^i{}_i) 
 +\left(\frac{\partial_0 r}{r}\right)^2\frac{r^2}{U^2}(\delta^{m'}{}_{m'}-2)
 \right] \gamma_{mn} , \\
 {\rm others} & =& 0 .
\end{eqnarray} 

\section{Solutions for vanishing and negative $\tilde{g}^2$} 

In this Appendix, we note the solutions for the case of vanishing or negative $\tilde{g}^2$. 
Following GGP, by introducing the variables
\begin{eqnarray}
&& x = \frac{1}{2}\phi + \ln A , \nonumber \\
&& y = \frac{1}{2}\phi+4 \ln W +\ln A, \\
&& z = -\phi -2\ln W, \nonumber
\end{eqnarray}
we obtain
\begin{eqnarray}
&& \left(\frac{d y}{d \eta}\right)^2 + 4 g^2 e^{2y} = \lambda_2^2 ,
\end{eqnarray}
and similar equations for $x$ and $z$, both of which are decoupled from $y$~\cite{Gibbons:2003di}.
Here, $\lambda_2$ is a constant for the first integral and
 $\eta$ is a coordinate in the coordinate system
\begin{equation}
ds^2_2 = W^8 A^2 d\eta^2+A^2d\psi^2 .
\end{equation}
The solution for a positive $g^2$ is presented in Ref.~\cite{Gibbons:2003di} as
\begin{equation}
y =
 -\ln \cosh(\lambda_2 (\eta-\eta_2))
  +\frac{1}{2}\ln\left(\frac{\lambda_2^2}{(4g^2)}\right) .
\end{equation}

However, as one can see, in Sec. IV B,
the effective $g^2$, namely $\tilde{g}^2$, can be positive or negative
in some time dependent solutions.
This then means that the solution for $y$ is replaced with
\begin{equation}
y = \lambda_2 (\eta-\eta_2) ,
\end{equation}
for the case of vanishing $\tilde{g}^2$ and
\begin{equation}
y =
 -\ln \sinh(\lambda_2 (\eta-\eta_2))
  +\frac{1}{2}\ln\left(\frac{\lambda_2^2}{(-4\tilde{g}^2)}\right) ,
\end{equation}
for a negative $\tilde{g}^2$. 

We then obtain
\begin{equation}
AW^4 =
 \left\{
\begin{array}{lll}
 \frac{1}{(\cosh^3(\lambda_1 (\eta-\eta_1))\cosh(\lambda_2 (\eta-\eta_2)))^{1/4}}
 \frac{\cosh(\lambda_1 (\eta-\eta_1))}{\cosh(\lambda_2 (\eta-\eta_2))}
 \left(\frac{\lambda_2^2}{4 \tilde{g}^2}\right)
 \left(\frac{q^2}{2\lambda_1^2}\right)^{-1}
 & {\rm for \,\,\,\,\, positive} & \tilde{g}^2 \\
 \frac{1}{(\cosh^3(\lambda_1 (\eta-\eta_1))\sinh(\lambda_2 (\eta-\eta_2)))^{1/4}}
 \frac{\cosh(\lambda_1 (\eta-\eta_1))}{\sinh(\lambda_2(\eta-\eta_2))}
 \left(\frac{\lambda_2^2}{-4\tilde{g}^2}\right)
 \left(\frac{q^2}{2\lambda_1^2}\right)^{-1}
 &{\rm for \,\,\, \, negative} & \tilde{g}^2 \\
 \frac{1}{(\cosh^3(\lambda_1 (\eta-\eta_1))e^{-\lambda_2 (\eta-\eta_2)})^{1/4}}
 \frac{\cosh(\lambda_1 (\eta-\eta_1))}{e^{\lambda_2 (\eta-\eta_2)}}
 \left(\frac{q^2}{2\lambda_1^2}\right)^{-1}
 &{\rm for \, vanishing}&  \tilde{g}^2
\end{array}
\quad .
\right.
\end{equation}
and 

\begin{equation}
A =
 \left\{
\begin{array}{lll}
 \frac{1}{(\cosh^3(\lambda_1 (\eta-\eta_1))\cosh(\lambda_2 (\eta-\eta_2)))^{1/4}}
 \left(\frac{\lambda_2^2}{4 \tilde{g}^2}\right)^{1/2}
 \left(\frac{q^2}{2\lambda_1^2}\right)^{-3/2}
 & {\rm for \,\,\,\, \, positive} & \tilde{g}^2 \\
 \frac{1}{(\cosh^3(\lambda_1 (\eta-\eta_1))\sinh(\lambda_2 (\eta-\eta_2)))^{1/4}}
 \left(\frac{\lambda_2^2}{-4\tilde{g}^2}\right)^{1/2}\left(\frac{q^2}{2\lambda_1^2}\right)^{-3/2}
 &{\rm for \,\,\,\,  negative} & \tilde{g}^2 \\
 \frac{1}{(\cosh^3(\lambda_1 (\eta-\eta_1))e^{-\lambda_2 (\eta-\eta_2)})^{1/4}}
 \left(\frac{q^2}{2\lambda_1^2}\right)^{-3/2}
 &{\rm for \, vanishing}&  \tilde{g}^2
\end{array}
\quad .
\right.
\end{equation}
Here, $\lambda_1$ is a constant for the first integral with repect with $x$, 
 and we used the solution of $x$ given in~\cite{Gibbons:2003di}.
Setting $\lambda_1 = \lambda_2 =1$ and 
 introducing a new coordinate $d r = AW^4d \eta$,
allows us to derive the deficit angle given in Eqs.~(\ref{defpos}), 
(\ref{defzero}) and (\ref{defneg}).



\end{document}